# Smart Close-out Netting

Akber Datoo[†] and Christopher D. Clack[‡]

October 2020

[†]D2 Legal Technology, Level 39, One Canada Square, Canary Wharf, London E14 5AB
akber@d2legaltech.com

[‡]Centre for Blockchain Technologies, Department of Computer Science, UCL, London WC1E 6BT
clack@cs.ucl.ac.uk, [1]

**Abstract**

Smart Close-out Netting aims to standardise and automate specific operational aspects of the legal and regulatory processes of close-out netting for prudentially regulated financial institutions. This article provides a review, analysis and perspective of these operational processes, their benefits for prudentially regulated trading institutions, their current inefficiencies, and the extent to which they are amenable to standardisation and automation. The main concepts of Smart Close-out Netting are introduced, including the use of a controlled natural language in legal opinions and the use of a data-driven framework during netting determination.

**Keywords:** Automation; Close out netting; Legal opinions; Regulatory capital; Risk; Standardisation

## 1. Introduction

Prudentially regulated financial institutions are required to manage the risk of their trading exposures and set aside regulatory capital in respect of exposure they have against their trading counterparties. This aims to manage the amount of capital at risk, a "buffer amount" that is held and available in the event of a counterparty's insolvency: an institution may have thousands of trades with each individual counterparty, each of which may create a positive or negative exposure. Despite the fact that these exposures may offset one another, in practical terms insolvency administrators might seek to characterise each such trade as a separate "contract" (notwithstanding that the trades together may be governed by a single, overarching master agreement), picking and choosing between the trades to be performed in insolvency ("cherry-picking"). Accordingly, the starting point under prudential banking regulations is that a financial institution must set aside capital in respect of the gross rather than net treatment of the exposures. Close-out netting seeks to mitigate the risk that a counterparty's insolvency administrator would claim selective performance of contractual trades profitable to the insolvent debtor and seek to repudiate the unprofitable contractual trades.[2]

Managing regulatory capital on a gross basis as described above is both onerous and expensive. However, in many jurisdictions (e.g. England and Wales), insolvency laws and broader regulations allow firms to view such exposures on a net rather than gross basis.[3] Typically this is permitted by regulators for regulatory capital purposes if an institution can demonstrate that they have obtained an up to date written and reasoned legal opinion that confirms the enforceability of "close-out netting" (see below) on an insolvency in respect of (a) the type of master trading agreement; (b) the type of counterparty; (c) the counterparty's jurisdiction; and (d) the type of transaction(s).[4]

---

[1] During the course of this work the second author provided consultancy services to D2 Legal Technology.

[2] Wood, P. R. (2007). *Set-off and netting, derivatives, clearing systems* (Vol. 4). Sweet & Maxwell, para 1-029 – 1-031.

Close-out netting is the contractually agreed cancellation of a series of open executory contracts (e.g. on-going trading relationships, rather than debts) between two parties on the default of one of the parties and the calculation of any close-out amount payable in respect of the cancelled future exposures by the netting of the resulting hypothetical gains and losses with reference to market prices or values. In the context of a master agreement, this allows the exposure under a bilateral trading arrangement to be viewed on a net rather than gross basis. To ensure legal enforceability of this arrangement, the parties' contractual netting agreement will seek to confirm creation of a single legal obligation, covering all included transactions such that there is a single net sum owing (that agreement being typically referred to as a "master agreement", due to this characteristic).

Very substantial amounts of regulatory capital therefore may be saved through obtaining positive legal opinions on the enforceability of close-out netting, and, without the resulting cost-efficiencies of close-out netting, it is unlikely that the swaps and securities trading markets would have reached anything like their current size and liquidity. For example, the Bank for International Settlements reports that as of the end of 2019 the effect of close-out netting for Over The Counter ("**OTC**") derivatives reduced gross exposures by 80% (from $11.6 trillion to $2.4 trillion) and that without close-out netting the trading institutions worldwide might therefore face a capital shortfall of over $9 trillion.[5]

Such successes in reducing gross exposures have led to industry bodies, such as the International Swaps and Derivatives Association (ISDA), providing assistance in the development of close-out netting legislation in respect of different jurisdictions. ISDA found that an increasing number of jurisdictions were seeking guidance on how to implement a comprehensive regime in support of close-out netting and related financial collateral arrangements in order to increase the safety and competitiveness of their domestic financial markets. ISDA reacted by developing the ISDA Model Netting Act which is a model law intended to set out the basic principles required to ensure the availability and enforceability of bi-lateral and multibranch close-out netting. The 2018 Model Netting Act draws on ISDA's thirty years of experience with policymakers and regulators across the globe to provide practical and useable netting legislation to increase legal certainty.[6] This has expanded the number of nettable jurisdictions with the result that the production, usage and maintenance of such close-out netting legal opinions has become increasingly prevalent, and is likely to continue in the future.

---

[3] The International Swaps and Derivatives Association (ISDA) provides a summary of specific netting legislation across the world at https://www.isda.org/2018/06/12/status-of-netting-legislation/.

[4] For example, the Qatar Financial Centre has a specific netting law, which was modelled after the "Model Netting Act 2006" published by ISDA. Canada in contrast relies on banking legislation which allows for automatic stays in insolvency to prohibit termination of an obligation to pay merely because a party has gone insolvent.

[5] BIS (2020) *Statistical release: OTC derivatives statistics at end-December 2019*. Bank for International Settlements. https://www.bis.org/publ/otc_hy2005.pdf.

[6] ISDA was incorporated in 1985. After the publication of the 1987 ISDA Interest Rate and Currency Exchange Agreement ISDA became more active in supporting close-out netting legislation for jurisdictions.

The processes for securing legal opinions for close-out netting,[7] for making a netting determination (utilising the legal opinion's analysis with regards to the relevant elements of the agreement, to the facts relating to a counterparty and to the transactions being carried out under such agreement), and for managing those opinions, are currently inefficient and costly (see Section 3.2). Here, we review and analyse these processes and the extent to which they are amenable to standardisation and automation. We analyse the structure of legal opinions on close-out netting and we introduce the concepts of Smart Close-out Netting including the use of a controlled natural language for legal opinions and a data-driven framework for netting determination.

## 2. Background

This section provides a brief overview of the basic commercial and legal concepts relating to close-out netting, the benefits with regards to regulatory capital and credit exposure, and the importance of securing legal opinions on the enforceability of close-out netting provisions in legal agreements.

### 2.1. Overview of close-out netting

One of the main considerations for prudentially regulated financial firms when trading is the efficient and optimised use of the balance sheet and regulatory capital. Regulatory capital is the capital such an institution is required to set aside by its financial regulator (usually expressed as a capital adequacy ratio of equity as a percentage of risk-weighted assets) to ensure that these institutions do not take on excess leverage and become insolvent (which, given the role of financial institutions, may more generally impact both confidence in the banking system and overall financial stability) thereby posing a more direct and particular risk for any potential lender of last resort. Capital requirements govern the ratio of equity to debt, recorded on the liabilities and equity side of an institution's balance sheet.

Given that trading exposures under trading arrangements are a major item in determining the regulatory capital amount to be set aside, close-out netting provides the opportunity to reduce a trading firm's regulatory capital requirement dramatically: globally, by about $9 trillion (see above) and for an individual bank with a gross exposure to OTC derivatives of (say) $10 billion, this would reduce exposure by about $8 billion. Moreover, the required amount of regulatory capital limits the credit an institution can make available as part of its business activities. Its optimisation is therefore of critical importance.

The ability to close-out on a net basis (i.e. for close-out netting to be available) for regulatory capital calculations however cannot be assumed. Regulation typically places a number of requirements and conditions that need to be met by a prudentially regulated firm to treat exposures on a net rather than on a gross basis. For example, in the EU, under Clause 2 of Article 296 of the Capital Requirements Regulation (CRR), two conditions must be fulfilled:

1. A single legal agreement controlling multiple transactions (i.e. a master agreement):
    *"a) the institution has concluded a contractual netting agreement with its counterparty which creates a single legal obligation, covering all included transactions, such that, in the event of default by the counterparty it would be entitled to receive or obliged to pay only the net sum of the positive and negative mark-to-market values of included individual transactions"*;
2. Written legal opinions:

---

[7] Netting is sometimes also called "offsetting".

> *"b) the institution has made available to the competent authorities written and reasoned legal opinions to the effect that, in the event of a legal challenge of the netting agreement, the institution's claims and obligations would not exceed those referred to in point (a). ..."*

Clause 2(b) then continues with a requirement for the legal opinion to give clarification on applicable jurisdictions, as follows:

> *"…The legal opinion shall refer to the applicable law:*
> *i. the jurisdiction in which the counterparty is incorporated;*
> *ii. if a branch of an undertaking is involved, which is located in a country other than that where the undertaking is incorporated, the jurisdiction in which the branch is located;*
> *iii. the jurisdiction whose law governs the individual transactions included in the netting agreement;*
> *iv. the jurisdiction whose law governs any contract or agreement necessary to effect the contractual netting"*

In effect, this creates a requirement to apply a series of tests for a particular trading relationship, in order to assess whether one can apply close-out netting for regulatory capital purposes or not. The relevant factors for these tests are contained within the "written and reasoned legal opinions" requirement and are strongly dependent on a number of specific factors, such as:

- the type of trading agreement and its governing law (e.g. the particular ISDA Master Agreement or Global Master Repurchase Agreement, and its governing law);
- specific terms negotiated between the parties;
- where the counterparty is incorporated and, if different, the location of the branch through which the counterparty is transacting (since this affects the relevant jurisdictions applicable to the insolvency analysis);
- the counterparty type (since insolvency analysis may differ – for example, according to counterparty location, insolvency treatment of a pension fund may differ from insolvency treatment of a trading company); and
- various trade facts (such as where collateral, if any, is located).

**2.2. Basics of close out netting in law**

This section briefly reviews the basic commercial and legal concepts regarding close-out netting which straddles matters of contract and insolvency law. It should be noted that the specific analysis will differ based on a number of factors, such as the governing law of an agreement and the place of incorporation of a counterparty to a master trading agreement.

**2.2.1. Transaction value**

The concept of close-out netting is based on the assumption that each financial transaction governed by a master agreement between the parties has a value which can be ascertained as of any particular point in time and brought into the calculation of a single net balance of what would be owing to or due from the counterparty if the counterparty were then to default.[8] For traditional financial contracts such as bonds and loans this is straightforward, but valuing derivatives or repurchase transactions can be more complex since they often impose obligations on both parties (hence, it is necessary to consider the balance between the values of the rights each party has under the contract and, where there is reliance on quotations for replacement trades, the creditworthiness of the determining party). In the OTC derivatives market, transactions are valued using a process known as "mark-to-market"; this defines the net value of all future payments and/or deliveries involved in respect of that transaction, and is the amount that would have to be paid, or which would be received, by a party if the transaction were terminated and replaced with an equivalent transaction, involving identical rights and obligations for the period remaining outstanding at current market rates with that party.[9] [10] Consequently, mark-to-market balances the present value (positive or negative in each case and taking into account relevant credit considerations and support) of all future obligations of both parties under the contract, and such future obligations will be subject to a number of market variables.[11]

---

[8] We will ignore the additional considerations that collateral arrangements might add to this, but essentially recourse to credit support, if any, should also be taken into account.

For example, the mark-to-market value of a "fair deal" interest rate swap should in theory be zero at the time of its entry, and thereafter will fluctuate according to market conditions such as the value movement of variables on the underlying and according to the number of payments remaining and the time to maturity of each (which reduces market exposure as the number and duration of these reduce over time).

### 2.2.2. Netting example

Consider a scenario where Bank A has multiple trades (transactions) with Bank B (see Figure 1). Some trades may be of positive value to Bank A ("in-the-money" – Bank A would need to pay a third party to replace Bank B on the other side), whereas the others are negative in value to Bank A ("out-of-the-money" – a third party would pay Bank A to replace Bank B on the other side). If Bank B is not performing (e.g. is failing or unable to make payments due under the various trades), then Bank A will usually want to replace them.[12] Typically, Bank A will want to close out all transactions with Bank B and make a claim on Bank B's estate for what is owed. There are two contrasting situations here:

1) *Net Close-out:* If Bank A can make a single net claim for the value of the whole portfolio of trades it has with Bank B, then the counterparty exposure is as low as possible, given the circumstances because of offsetting positions. Transactions with a positive value are balanced against those with a negative value, resulting in a final net value.
2) *Gross Close-out:* If Bank A **cannot** make a net claim (i.e. if close-out netting is not possible) – then Bank A must:
   a) pay Bank B the full value of each transaction where Bank A owes money to Bank B; and
   b) make a legal claim against Bank B for each transaction under which it has credit exposure to Bank B (i.e. where Bank B owes money to Bank A), recognising that Bank B's estate may be insufficient to meet the full value of all such claims.

The gross close-out of the second situation increases the counterparty credit risk. Figure 1 illustrates an example for just three trades, from the perspective of Bank A: two of these are "in the money" and one is "out of the money". If we consider the possibility of the insolvency of Bank B, there are two scenarios:

1) *Net Close-out.* If we could net the mark-to-market values of the transactions, the value of the third would be subtracted from the sum of the values of the first two: overall this portfolio has a net value of -£100m from Bank A's perspective (Bank A owes £100m to Bank B).

    - This represents a credit exposure of £100m from Bank B's perspective, which must be managed. If Bank B were prudentially regulated, it would be required to set regulatory capital against just this net exposure.

2) *Gross Close-out.* If close-out netting were not possible then (i) Bank A would pay Bank B the mark-to-market value of each transaction where Bank A would owe Bank B money (i.e. the full total of £500m for Transaction 3), and (ii) Bank A would make a legal claim on the estate of Bank B for each transaction where the trade is in-the-money to Bank A (i.e. a total of £400m).

---

[9] Datoo, A. (2019). *Legal Data for Banking: Business Optimisation and Regulatory Compliance*. John Wiley & Sons.

[10] Hudson, A. (2013). *The Law of Finance* (2nd edn), Sweet & Maxwell, para 2-12.

[11] Datoo (op. cit. Footnote 9) comments: "*The mark-to-market value of a transaction is (typically) closely related to the replacement cost, but practically there are a number of nuances to this, such as bid/offer spreads, which may be significant in respect of illiquid product types. Additionally, replacement costs will naturally contain various valuation adjustments (known as XVAs), such as CVAs (credit value adjustments)*".

[12] A principle reason for wanting to replace them is that Bank A will want to ensure its various hedging arrangements (created by other transactions) remain balanced despite Bank B's non-performance.

The success of the legal claim is not certain – as an unsecured creditor, recovery might be only a penny in the pound in which case a claim for £400m might only receive £4m.[13] Since the legal claim might fail entirely, this represents a credit exposure of £400m from Bank A's perspective. Now Bank A, if prudentially regulated, must set aside regulatory capital against this full gross exposure.[14]

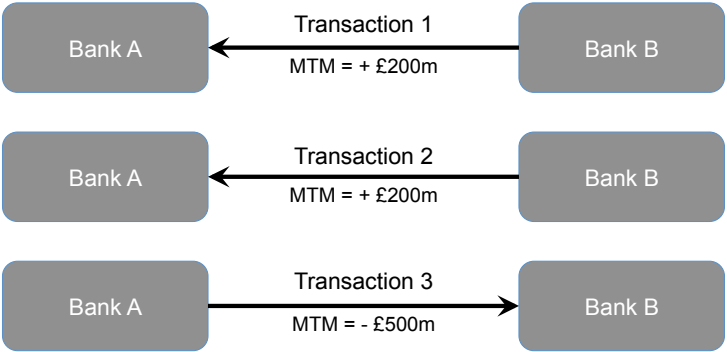

Figure 1: Example of mark-to-market values from the perspective of Bank A based on a portfolio of three transactions with Bank B. Transactions 1 and 2 are "in-the-money" whereas Transaction 3 is "out-of-the-money". With a net close-out, only Bank B has a net credit exposure (of £100m), whereas a gross close-out leads to credit exposures of £400m for Bank A and £500m for Bank B respectively.

---

[13] An insolvency practitioner might "cherry-pick" and entirely disclaim all unprofitable transactions, while enforcing the profitable transactions.

[14] Note that if Bank A were to become insolvent then Bank B would have a gross credit exposure of £500m against which, if prudentially regulated, it must set regulatory capital.

### 2.2.3. Legal opinion

A net close-out is desirable as a scenario that reduces counterparty credit risk and regulatory capital costs but may be subject to legal risk: it may be challenged due to possible issues with its enforceability under some insolvency laws. Jurisdictions have different stances towards close-out netting arrangements – a jurisdiction may be more pro-debtor or alternatively more pro-creditor in the orientation of its legal regime – and the insolvency laws in many jurisdictions expressly recognise close-out netting, but other jurisdictions might be more likely to allow insolvency administrators to cherry-pick between the contracts to be performed in an insolvency. Indeed, the insolvency laws in many jurisdictions dictate that the insolvency practitioner's duty is to assess what recoveries can be made and prevent unwarranted claims by the creditors to the insolvent party. As such, a liquidator who fails to attempt cherry picking when possible might be regarded as being negligent in the carrying out of their duties.

A master agreement declares the intentions of the parties to treat a large number of trades (transactions) as aspects of a single whole; thus, each new transaction merely varies or supplements the terms of the existing single agreement to include the cash flows of the new transaction. However, there is no guarantee of the enforceability of viewing the matter in this way: the key issue is whether a court would respect this arrangement of a net close-out or if they would seek to impose gross close-out and thus open the door to the possibility of cherry-picking.

This is the subject matter of a close-out netting legal opinion – opining on how a court would treat an agreement in this regard upon the insolvency of the parties thereto. As discussed above in Section 2.1, the ability to gain the benefit of close-out netting for regulatory capital purposes is subject to the requirement that close-out legal opinions be obtained in advance, to provide sufficient comfort on the enforceability of the close-out netting provisions. A close-out netting legal opinion will, as well as dealing with legal concerns and risks applicable to close-out netting, in some cases cover wider issues raised for example in terms of political or other jurisdictional risks. The view expressed might therefore be that close-out netting should be effective but nonetheless other circumstances do not eliminate cherry picking as a risk.

### 2.3. Standardisation and industry Master Agreements

Historically, parties to OTC derivatives and other capital markets transactions would document each and every transaction under a separate agreement. These often became quite lengthy documents, setting out not only the economic terms of the trade between the parties, but also terms relating to the general legal and credit relationship between the parties – such as termination rights in the event one of the parties failed to make payment or deliveries required under the transaction, or upon the insolvency of one of the parties. These would be repeated each and every time new transactions were entered into between the parties. To address this issue, a framework master agreement was created by various trade associations in respect of certain product types: for example, by ISDA for OTC derivatives, by the International Capital Markets Association[15] (ICMA) for repurchase transactions, and by the International Securities Lending Association[16] (ISLA) for securities lending transactions.

Aside from making the documentation process more streamlined and efficient, this also enhanced the possibility of close-out netting, which is now often said to be the primary purpose of a master trading agreement. By way of example, the ISDA Master Agreement standardised form (the "preprint", which is then negotiated and amended by trading parties to fit their commercial relationship), contains a number of elements designed to try and assist with the likelihood of a jurisdiction recognising the close-out provisions it contains as being enforceable. These are:

1. *The Single Agreement Provision*

---

[15] Formed in July 2005 from the merger of International Primary Market Association and the International Securities Market Association (formerly the Association of International Bond Dealers).

[16] Formed in 1989 to represent the common interests of participants in the securities lending industry.

    By its terms, each ISDA Master Agreement (inclusive of the preprint, Schedule and any annexes such as the Credit Support Annex to collateralise the trading relationship), together with any trade confirmation, including any relevant definitional booklets applied to the trade, entered into between the two parties, forms a single agreement.

    The concept of these all forming a single agreement, and not a series of agreements, evidences the intention of the parties and reinforces the argument that a liquidator or insolvency practitioner cannot break the 'single agreement' in order to cherry-pick profitable transactions without thwarting that intention.[17] [18]

2. *Flawed Asset Provisions*

    A flawed asset provision seeks to 'flaw' an asset (in the case of the ISDA Master Agreement, the obligation to make payments and/or deliveries by a non-defaulting party under transactions to the ISDA Master Agreement) without terminating the transactions and, therefore, avoid the step of crystallising a mark-to-market process. This is done by adding a number of conditions precedent before any future obligation to make payments and deliveries becomes a debt owed to any estate.[19]

3. *Close-out Mechanics*

    This generally involves the termination of all transactions on the insolvency or default of one of the parties to the agreement, and the contractual "liquidation" of any damages owed instead, by aggregating the values of each of those transactions and replacing the transaction payment amounts with a single sum representing all transaction payment amounts.[20] By replacing the various payments with a single amount, this again assists with the argument against an insolvency practitioner or liquidator seeking to cherry-pick certain transactions, since the individual transactions no longer exist, having been replaced with a single close-out settlement payment.

The master agreement preprint forms published by trade associations such as ISDA and ICMA greatly assist with the feasibility of meeting the prudential regulatory requirements to be able to apply close-out netting for the purposes of regulatory capital calculations and holdings. Without these standard forms, a prudentially regulated bank would require a written and reasoned legal opinion for each and every master trading agreement entered into. Where a large investment bank might hold several tens of thousands of active master trading agreements, it would not be feasible or cost-efficient to obtain separate close-out netting opinions in respect of each one, nor to update these opinions regularly as required by the regulators.

---

[17] This single agreement concept is contained in section 1(c) of the 2002 ISDA Master Agreement.

[18] Drawing from the reflections of Jeffrey Golden, one of the architects of the close-out netting provision of the ISDA Master Agreement, on his research in this area the inference from many cases that he looked at suggested cherry-picking was a construct designed to recognise, where appropriate, the parties' intention that two relationships should be treated as separate.

[19] In the case of a 2002 ISDA Master Agreement, the flawed asset provision is contained in section 2(a)(iii). Section 6(c)(ii) should be noted as the flip-side of this, with section 2(a)(iii) providing the condition and section 6(c)(ii) operating to suspend performance upon the occurrence of the condition. This provision under the laws of England and Wales was considered by the courts in Lomas v JFB Firth Rixson, Inc. [2012] EWCA Civ 419.

[20] The close-out mechanics are contained in Section 6 of the 2002 ISDA Master Agreement.

The use of the standardised form agreement preprints also means that prudentially regulated firms can instead rely on, at least as a starting point, industry-sponsored close-out netting legal opinions that cover any agreement based on and negotiated from the starting point of such industry forms, provided no material changes have been made by the parties that might disturb the close-out netting relevant provisions in such a way as to affect the analysis on which the opinion relies, thereby dramatically reducing the number of close-out netting legal opinions required to be obtained and maintained.

## 3. Analysis

The advantages of close-out netting for a prudentially regulated institution have been outlined in the previous section. This section analyses the current processes involved in obtaining and using legal opinions for close-out netting, including an analysis of the typical structure of a legal opinion and an assessment of some of the sources of inefficiency in current practice.

### 3.1. Lifecycle processes

Figure 2 illustrates the key processes and data objects involved in obtaining and using legal opinions for close-out netting. These processes are triggered by the desire to trade on a commercially favourable basis, and especially to obtain a benefit with regards to the trading institution's regulatory capital position.

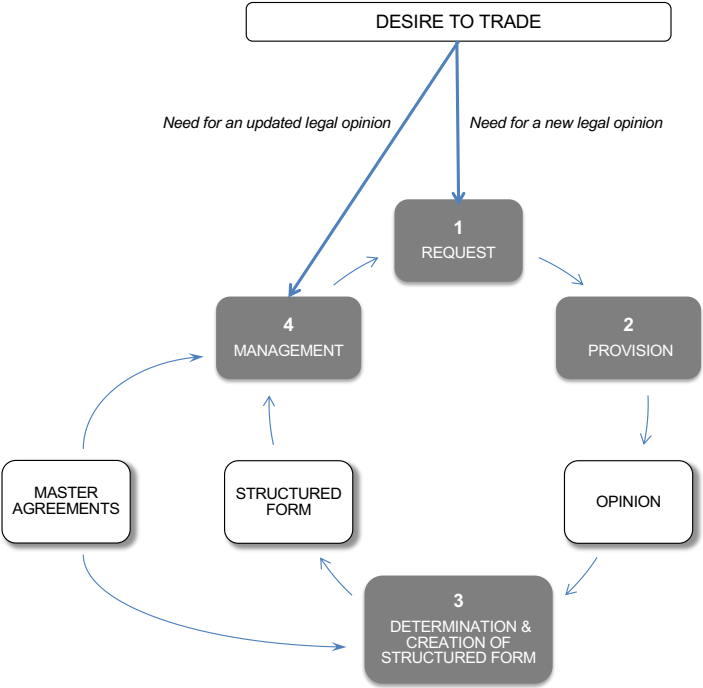

Figure 2: Obtaining and using legal opinions for close-out netting. Processes are dark grey boxes: data objects are shadowed white boxes.

The four main processes are explained below and the key data object (the legal opinion) is further analysed in Section 3.1.

1. Request for opinion.

The trading institution requests a law firm to provide a specific legal opinion on close-out netting – this may be an initial base opinion or may be a request for an update to a previously requested opinion. The trading institution will include specific questions to be answered and will provide key data that include the following:

- the type of agreement (e.g. the particular ISDA Master Agreement or Global Master Repurchase Agreement) and its governing law;
- the relevant jurisdiction(s) applicable to the insolvency analysis (mainly centred on the jurisdiction of incorporation of the counterparty – and any relevant branch, if any – and the location of the collateral);
- the counterparty type (this is driven by the fact that the insolvency analysis may be different for different counterparty forms of organisation, for example an insurance company compared to a corporate, including various public policy considerations when one considers counterparty types such as local authorities);[21] and
- other jurisdiction-specific variables such as the specific transaction types covered by the agreement – for example, some derivative transactions may only comply with the relevant netting legislation in a given jurisdiction if they meet certain conditions.

2. Provision of opinion.

The law firm provides a written legal opinion, which may be a new opinion or a revision of a previous opinion (e.g. to give regard to changes in the law or changes in the facts since the previous opinion, to change the legal analysis, and/or to add or remove certain qualifications and scope such as counterparty type).

3. Netting determination.

For each trading relationship (master agreement), the trading institution carefully reviews the relevant legal opinion and how it applies to the fact pattern at hand in terms of counterparty, legal agreement and trade data. Increasingly, regulators expect firms to analyse each assumption and qualification contained in a legal opinion, considering whether any might affect the end conclusion as to whether close-out netting would apply for regulatory capital purposes or not. There is an expectation of applying a similar approach for each equivalent assumption and/or qualification across a range of obtained legal opinions, and thus leading to the creation of databases of these assumptions and qualifications. This in turn, where well documented, greatly assists with the audit trail and data lineage requirements behind a netting determination – and, for major financial institutions, assists compliance with the principles of the Basel Committee on Banking Supervision's Principles for effective risk data aggregation and risk reporting.[22] [23] These BIS principles increase the need for data governance and, in order to make

---

[21] Additionally, an institution may include whether its intended counterparty wishes to include branches.

[22] BIS (2013) 'Principles for effective risk data aggregation and risk reporting', *Bank for International Settlements*. BIS Basel Committee on Banking Supervision. https://www.bis.org/publ/bcbs239.pdf.

[23] The objective of BCBS 239 (op. cit. Footnote 22) is to ensure systemically important financial institutions' risk data aggregation capabilities and internal risk reporting practices are sufficient and to enhance risk management and decision-making processes. Several of the key principles under it would be supported by increased automation

the process practically manageable and controlled from an operational risk perspective, increase the need for automation in this area.

In summary, the trading institution:

- determines whether the legal opinion is positive or negative for a given trading relationship; and

- effectively converts the legal opinion and relevant facts (agreement, client and trade-related) into an internal structured summary form and then into a binary "Yes/No" flag for use within the management of trade data, client data and legal agreements.

4. On-going management.

The trading institution manages a large collection of legal opinions and netting determinations – this is greatly assisted by holding the opinions in a structured form, and includes (i) managing the application of legal opinions to client agreements, (ii) managing the setting aside of regulatory capital, (iii) managing the assumptions and qualifications contained in those legal opinions, and the consequential need for updates and alerts which would trigger a repeat request for a legal opinion. There are several possible reasons for seeking an updated opinion, for example:

- changes not contemplated by the original opinion may have been made to the legal agreement for commercial, regulatory or operational reasons;

- the legal reasoning changes over time (e.g. due to new regulation or statutory change, new case law, changing best practices expectation) and so the opinion must be regularly refreshed;

- changes may have been made to the trades (including to their mark-to-market, termination and/or addition of trades), which may impact the value of the net exposure;

- extreme events (such as Brexit or COVID-19) causing a reassessment of contractual liabilities;

- update to a pre-issued legal opinion (including adding or removing certain qualifications and/or scope such as counterparty type) may change the legal analysis / rules and require other opinions to be updated; or

- the passing of time – many firms decide on a set period for which the conclusions on a legal opinion may be relied upon before needing to be renewed (subject to manual override and exceptions) e.g. 12 months from the date of issuance, after which the conclusion is set to "No" if the opinion is not renewed (this would then be viewed as an "update to a pre-issued legal opinion").[24]

---

and data standards in relation to the entire close-out netting process: for example Principle 2 (Data architecture and IT infrastructure), Principle 3 (Accuracy and Integrity), Principle 5 (Timeliness) and Principle 8 (Comprehensiveness).

[24] Note the requirement in the CRR (Article 297 Obligations of institutions para. 1.). *"An institution shall establish and maintain procedures to ensure that the legal validity and enforceability of its contractual netting is reviewed in the light of changes in the law of relevant jurisdictions…."*.

### 3.1. Legal opinion structure

A close-out netting legal opinion will typically comprise five parts: scope, assumptions, qualifications, discussion/analysis of the issues and conclusion.[25]

### 3.1.1. Scope

A legal opinion will consider the validity and enforceability under a particular governing law (e.g. the laws of England and Wales) of transactions under particular forms of a master trading agreement (e.g. the 2002 ISDA Master Agreement or 1992 ISDA Master Agreement (Multi-Currency – Cross Border)).

The scope will specify any particular transaction types the opinion might or might not cover. Insolvency rules, as well as other considerations on close-out netting, might differ based on the type of counterparty being faced under the agreement. The scope of the legal opinion might therefore, for example, be limited to particular counterparty types (e.g. a corporation registered as a company in England and Wales under Section 1 of the Companies Act 2006).

### 3.1.2. Assumptions

The law firm giving an opinion will have to assume that a certain fact pattern applies in order for it to reach the conclusions stated in the opinion. Thus, most opinions will contain an identifiable list of assumptions that set out the statements of fact on which counsel has relied for the purposes of giving the opinion.

Assumptions may for example be factual (e.g. that *"The Master Agreement and each Transaction were entered into by both parties to the master trading agreement prior to an insolvency type event"*) or may be related to the legal agreement (e.g. that *"No provision of the ISDA Master Agreement that is necessary for the giving of our advice in this memorandum has been altered in any material respect"*).[26]

Assumptions may be generic, or specific. For example:

- A generic assumption regarding incorporation: *"The counterparty is duly organised and validly existing in good standing under the laws of its relevant jurisdiction and its place of business"*. (This assumption may be verified by legal due diligence by reference to publicly available information on the good standing of the parties/or as provided by the counterparty).
- A specific assumption regarding the intentions of the parties: *"On the basis of the terms and conditions of the ISDA Master Agreement and other relevant factors, and acting in a manner consistent with the intentions stated in the ISDA Master Agreement, the Parties over time enter into a number of Transactions that are intended to be governed by the ISDA Master Agreement"*. (This assumption may be verified by each transaction being documented in the form of a confirmation referencing the relevant master agreement – and ensuring there is nothing to indicate the contrary).
- An assumption may also be stated as a condition precedent (e.g. assuming the parties *"have elected for the Automatic Early Termination provision of Section 6(a) of the ISDA Master Agreement"*).

---

[25] This structure is illustrated by Allen and Overy's *Validity and Enforceability under English Law of Close-out Netting under the 2002 and 1992 ISDA Master Agreements* published 12th March 2019.

[26] These examples are taken from the Swiss Memorandum of law on the enforceability of the termination, bilateral close-out netting and multibranch netting provisions of the ISDA Master Agreements, published by Lenz & Staehelin on 7 May 2019.

### 3.1.3. Qualifications

Unlike assumptions (which, broadly speaking, are statements of fact), qualifications are included in legal opinions in order to clarify the limitations of the conclusions reached as a matter of law. Netting opinions aside, for example, enforceability opinions are often qualified by reference to insolvency law, meaning that the addressee is advised not to rely on the conclusions in the scenario where one (or both) parties to the relevant agreements is insolvent or subject to bankruptcy/insolvency proceedings. Whilst, as a technical matter, assumptions and qualifications perform distinct functions, there is often considerable overlap in practice.

Examples of general qualifications include:

- Fraudulent contracts: "A party may be able to avoid its obligations under a contract where it has been induced to enter into it by a misrepresentation, bribe or other corrupt conduct. A court or arbitral tribunal will generally not enforce an obligation if there has been fraud."
    - This qualification would be verified by conducting thorough legal due diligence and ongoing compliance / anti-fraud controls against the counterparty.
- Scope of opinion: "The opinion is limited to the specific queries asked on the ISDA Master Agreement and should not be viewed as a general enforceability opinion under the relevant law with respect to the 2002 ISDA Master Agreement."
    - This limitation is driven by the scope of the questions that have been asked to be addressed in the opinion and goes to the enforceability of close-out netting provisions. A trading institution will typically conduct additional legal due diligence as required on a case-by-case basis in respect of all other matters.
- Jurisdiction: "courts may stay insolvency proceedings where they are of the opinion that proceedings in another forum would be more convenient or if concurrent proceedings are being brought elsewhere."
    - This qualification is verified by review of the governing law of the Agreement and ensuring it adheres to the relevant assumptions and qualifications.

### 3.1.4. Discussion / Analysis of the issues

This section generally sets out the insolvency rules applicable in the jurisdiction, and whether there is specific legislation protecting close-out netting of financial contracts. The opinion writer will usually analyse the close-out netting provisions in the relevant agreement type(s) in the light of the insolvency rules and any netting legislation in order to provide the reasoning for the conclusion as to what extent these provisions would be valid and enforceable in an insolvency.

### 3.1.5. Conclusion

Depending on the legal certainty in the jurisdiction, the conclusion(s) in the legal opinion may be worded with varying degrees of strength. Similarly, recommendations to the financial institution to whom the opinion is addressed may carry varying degrees of strength. For example:

- 'a court would…',
- 'a court may…',
- 'it is likely that a court would…',
- 'an institution can…'
- 'an institution should…'.

We observe that this inevitably raises a question regarding how to map nuanced natural language terms into a binary outcome, since the regulatory capital test is binary: for these purposes, an agreement can only be nettable or non-nettable. Consequently, due to the varied strength of certain conclusions in the legal opinion and other factors, an institution must conduct its own analysis of the legal opinion and draw its conclusions as to the robustness of an opinion in order to make this binary decision. It is the financial firm that must in the first instance determine that exposures either may or may not be considered net for regulatory capital purposes, based on the written and reasoned legal opinion.

### 3.2. Sources of inefficiency in current practice

Both initial netting determination and on-going management require assessment of the relevant master agreements[27] and the obtained textual legal opinions; sometimes repeated assessment of the same legal opinion is required, and the complexity and subtlety of the embedded assumptions and qualifications can make this costly. In many institutions that cost is mitigated to some extent by converting the legal opinion into a structured form that is easier and faster to assess during the management of trade data, client data and legal agreements. Nonetheless, in some cases even this structured form presents usability issues and does not satisfy the regulator's desire to see an institution analyse a legal opinion.

There is no agreed standard structured form of a legal netting opinion and the summary forms created by the recipients of such opinions therefore tend to differ between trading institutions. Although details differ, in essence there will be a recording of certain key data (such as the date of the legal opinion, the jurisdiction to which it relates, the agreement types it covers, and the law firm that provided the opinion), and perhaps some other fields for workflow processing, and then the entire legal text may be saved with added comments and annotations to highlight issues that require special consideration for the relevant institution. Example annotations include:

- whether the answer or analysis is generally positive, neutral, negative or missing (i.e. the legal opinion doesn't answer the question);
- issues requiring general consideration or significant issues in relation to regulatory compliance; or
- specific qualifications that require consideration for this institution.

The aim of these annotations is to capture a range of judgements made during the initial netting determination, so that effort is reduced during subsequent re-assessments that require re-reading of the legal opinion. The nature of the annotations should therefore anticipate the ways in which future management processes will need to re-assess the legal opinion.

However:

- where structured forms are maintained only as text this may not be easily understood by computer software; this therefore requires a human to read the text and reduces the opportunities for onward automation;
- as noted above, structured forms may differ between institutions, thereby reducing the opportunity for the development of industry-wide tools for automation; and

---

[27] In particular, the schedule for each master agreement (which is where many modifications are often made to the preprint). In some cases, confirmation documents for individual transactions may contain modifications and this may lead to a need for further assessment. The non-standard nature of such modifications is another source of inefficiency and cost.

- despite a collective need, each trading institution obtains and maintains legal opinions individually.

### 3.2.1. A simple cost model

The costs of netting determinations include:

- the actual costs associated with human assessment of master agreements and legal opinions, conversion of legal opinions into a structured form, and on-going management;
- lost opportunity cost in terms of "false negative" determinations that lead to excessive regulatory capital being set aside;
- the risk cost of "false positive" determinations that lead to too little regulatory capital being set aside;
- the imposition of regulatory sanctions or fines due to "false positive" determinations or poor management of the end to end close-out netting process (and therefore in the provision of regulatory capital); and
- the consequences of overlooking the need for a new or updated opinion.

The latter can arise due to poor data governance. For example, poor standards and lack of suitable taxonomies across the client, product and legal agreement data domains may lead to confusion over whether an opinion has been provided on the correct basis, or there may be confusion over the reasoning that justifies the opinion. Example problems include: (i) confusion over product type, e.g. where the institution's taxonomy differs from the law firm's taxonomy, (ii) confusion over counterparty type, e.g. where the financial institution uses a definition from its internal taxonomy created for Anti Money Laundering (AML) checks that may not map correctly to the taxonomy used by the law firm, and (iii) for larger institutions with multiple subsidiaries, confusion as to whether separate legal opinions are required for different subsidiaries. For reasons of cost, prudentially regulated financial institutions tend to obtain generic opinions for a particular agreement type, jurisdiction and counterparty type – and then make assumptions (e.g. one does not materially amend any positions in a standard form ISDA agreement that would disturb netting), that may not be correct.

As a concrete example, we provide a simple cost model to illustrate the scale of <u>just one</u> of the above costs, i.e. the cost of conversion of a legal opinion into structured form, expressed for the sector as a whole.

A typical large trading institution such as an investment bank may need to manage several hundred legal opinions. Each year, a certain percentage of these opinions need to be sent for legal review, and that percentage varies according to the extent of trading undertaken by an institution. A simple way to measure the extent of trading (for prudentially regulated institutions) is to use Tier-1 risk-based capital as a proxy and identify clusters – for example, available data regarding 5,265 United States banks in September 2019[28] can be arbitrarily sectioned into the following five levels:

---

[28] https://www.usbanklocations.com/bank-rank/tier-1-core-risk-based-capital.html.

- Level 1 - 20 banks with over $15bn of Tier-1 capital[29]
- Level 2 - 56 banks with $2.25bn - $15bn of Tier-1 capital
- Level 3 - 162 banks with $500m - $2.25bn of Tier-1 capital
- Level 4 - 649 banks with $100m - $500m of Tier-1 capital
- Level 5 - 4,378 banks with under $100m of Tier-1 capital

The lowest level (Level 5) comprises those banks that undertake only limited trading and will therefore manage far fewer legal opinions for close-out netting; we will not consider them further. As the size of the firm reduces from a Level 1 to Level 4, the volume and type of products traded will typically reduce, as will the jurisdictional footprint of the counterparties, and the combination of these two factors will lead to a non-linear positive correlation between size and the total number of opinion reviews required. Let us assume (an anecdotal estimate, from practical experience) that there are 300 legal opinions available, each providing a unique combination of jurisdiction, counterparty type and agreement type coverage. Although a Level 1 bank may rely on 80%[30] of these (and therefore need to review each, and any updates, typically annually), a Level 4 bank may only need to review (say) 5% of the available 300 legal opinions.

Each time a legal review is undertaken, the legal text must be converted afresh into a structured form. If the cost for converting each legal opinion were the same, and if we knew the annual percentages reviewed at each Level, we could use a simple formula to derive a rough estimate of the overall sector impact of the annual cost of this component of netting determination as follows. For illustration in this example we have used 80%, 40%, 10% and 5% as the legal opinion coverage percentages at Levels 1, 2, 3 and 4 respectively:

- Level 1 – 20 banks × (say) 300 opinions × (say) 80% reviewed each year = 4,800 reviews per annum
- Level 2 – 56 banks × (say) 300 opinions × (say) 40% reviewed each year = 6,720 reviews per annum
- Level 3 – 162 banks × (say) 300 opinions × (say) 10% reviewed each year = 4,860 reviews per annum
- Level 4 – 649 banks × (say) 300 opinions × (say) 5% reviewed each year = 9,735 reviews per annum

TOTAL: 26,115 reviews per annum

When considering the workload cost of these reviews, we observe that in practice legal opinions tend to be either complex or simple, and the conversion costs associated with the former are very much higher than the latter. We can therefore derive a simple model of the total annual sector cost in days "$TC_d$" as follows:

$$TC_d = \Sigma_L \ (banks_L \times opinions_L \times reviewed_L \times ((complex_L \times cost_C) + ((1 - complex_L) \times cost_S)))$$

where

$L \in \{1,2,3,4\}$ is a Level number

---

[29] Typically, these are also Global Systemically Important Banks: see for example https://www.fsb.org/wp-content/uploads/P221119-1.pdf.

[30] Note that if one took two Level 1 banks, although they may both rely on (and therefore review) some 240 opinions each, it is likely that each would manage a different set of 240 opinions, based on their sales and trading coverage.

$opinions_L$ is the total number of opinions managed by a bank at Level L

$complex_L$ is the percentage of complex opinions managed by a bank at Level L

$banks_L$ is the number of banks at Level L

$reviewed_L$ is the percentage of opinions reviewed each year by a bank at Level L

$cost_C$ and $cost_S$ are the costs of reviewing a complex or simple opinion, respectively, and comparing those opinions against the relevant master trading agreements.

Inserting some anecdotal numbers from practical experience will provide a rough numerical estimate of $TC_d$. For example: $opinions_L$ = 300 for all L, $complex_L$ = 0.5 for all L, $reviewed_L$ = 0.8, 0.4, 0.1 or 0.05 depending on L, $cost_C$ = 2 days and $cost_S$ = 0.25 days. This would give $TC_d$ = 29,379.38 days, about 37% of which arises from the Level 4 banks. At roughly £1,000 per day this model demonstrates a total sector cost in excess of £29m per annum. Furthermore, when these costs are viewed in the context of a failure to optimise the balance sheet opportunity, along with sanctions and fines being imposed due to mistakes in netting determinations or process control failures, this figure of £29m is likely to be far exceeded in practice.

The above model provides an initial estimate of conversion costs but makes the assumption that for a given combination of counterparty type, jurisdiction and agreement type all agreements are "standard" – which of course they are not. This fact will tend to further inflate the above annual costs for the sector, especially for the banks in Levels 1 and 2 (where there is most departure from "standard" agreements). Furthermore, the term $banks_L$ has been based on US data and so the estimated total sector cost of £29m per annum relates to the US only. We have argued previously that $banks_1$ and $banks_2$ are probably correct globally, but the numbers for Levels 3 and 4 will be underestimates for the global sector, and so total global sector costs will be higher.

## 4. Smart close-out netting

Smart close-out netting aims to use standardisation to remove key sources of operational risk in close-out netting determination. For example the risks associated with: (i) the translation from nuanced natural language into a binary "Yes/No" decision; (ii) variations between law firms in the use of natural language expressions, and the use of non-standard language by law firms; and (iii) the text of the legal opinion becoming separated from the derived data from that opinion (and subsequently not being available to resolve queries that might arrive from the use of the derived data in the context of a specific master agreement).

Specifically, smart close-out netting uses a controlled natural language for legal opinions and a standard framework within which to make netting determinations. The controlled natural language is:

- used by law firms to provide legal netting opinions; and then
- used internally within trading institutions within a standard framework:
  - to capture and store the details of a legal opinion in a standard way;

- to support standardisation and automation of the operational mechanisms for comparing legal opinions against the facts of individual master agreements, and for making and storing netting determinations;[31] and
- to take internal positions based on the external legal advice with reference to their own risk tolerance.

As discussed in Section 3.2, there are existing systems that represent legal opinions systematically in a logical structured format. For example, *netalytics*, a joint venture between ISDA and aosphere, provides a platform which analyses and represents legal opinions in a structured format, displaying the content of opinions consistently as data.[32] Colour-coded references are provided as to the likely strength of an opinion's assertions and outcomes, representing the data in a way that is easily understandable and structured. While institutions cannot ordinarily rely on the system for close-out netting, it is widely used to assist in-house reviews and summaries of close-out netting opinions.

Although the *netalytics* approach creates a structured data-driven view of the legal advice from the natural language legal opinion, the opinion remains separate from the structured form and there is a necessary interpretative exercise conducted on the legal opinion after it has been provided. This is a source of inefficiency and operational risk; it creates opportunities for misaligned understandings between the lawyers providing the legal opinion, and those creating these structured data summaries.

Further, this risk of misaligned understanding can be mapped within an institution to downstream consumers of the structured data summaries. The most beneficial option may be to remove the need for mappings by institutions and instead have a system which presents the legal opinion data in a structured format *ab initio*. This will greatly benefit an institutions risk and trade systems by reducing operational risk.

Smart close-out netting aims to remove this inefficiency and risk by the expedience of having the legal opinion expressed in a structured form (using a controlled natural language) from the outset by the law firm, creating opportunities to streamline the netting determination process and the use of data analytics in which to ground business decisions, taking risk tolerances into account.

A well-defined and governed data structure that represents the legal opinion can act as the basis for an institution to make its netting determination: adjusting as necessary the conclusion expressed by a law firm, based on weightings given to common and bespoke considerations and on additional facts relevant to the netting determination that the law firm would not have had available to it when providing the legal opinion itself. It provides a foundation upon which an institution's particular risk tolerance, risk analysis approach, and various factual matters individually relevant to it, can be calibrated.

### 4.1. Standardising legal opinions

In Section 3.1 a generic five-part structure for the legal opinion was provided: scope, assumptions, qualifications, discussion and conclusion. Unfortunately, not all legal opinions have clear delineations between the five sections, nor do they have a clearly and consistently defined standard structure (though there have been attempts by industry bodies to create more formal separation – for example by including an appendix on a summary of the recommended or required modifications of an agreement). [33]

---

[31] Full automation of these processes is not an aim, due to the requirement for the trading institution to apply judgement, but it is an aim to automate those processes that support the making of a judgement.

[32] https://www.aosphere.com/aos/netalytics.

### 4.1.1. The discussion

Of the five parts of a legal opinion, arguably the most difficult to standardise is the discussion. The purpose of the discussion is to demonstrate and substantiate that the law firm has come to a reasoned conclusion on whether something is nettable or non-nettable: it is also necessary in order for an institution to ascertain for itself that the conclusion is based on sound reasoning and therefore demonstrate that it has been prudent in accepting the logical conclusions based on the discussion. Specifically, institutions are required by Article 296(2)(b) of the Capital Requirements Regulation (CRR) to have "reasoned legal opinions", which has been interpreted (and confirmed by regulatory implementation) to mean that institutions must conduct an analysis of the opinions they seek to use. Institutions may, upon a review of the discussion, decide that the conclusion is not strong enough for them to support it, despite the fact the logic is correct.

The discussion is likely to be difficult to codify. Firstly, it will need to contain the factual basis of insolvency law in a jurisdiction: this necessitates an examination of complex legal concepts which, depending on the jurisdiction, will originate from judicial precedence, domestic code, domestic statute, and domestic and international regulation. These factual concepts are often interdependent, for example a domestic statute concerning international regulation; therefore, codifying these interdependencies would be complex. Secondly, the factual basis of insolvency is overlaid by the law firm's analysis of the application of insolvency law to the covered entities of the opinion. The factual basis and analysis on the application of insolvency law in a given scenario will involve complex natural language. Hence, without significant manual oversight there is a risk any codification may not represent the analysis in its entirety.

### 4.1.2. The conclusion

By contrast, the conclusion is likely to be more amenable to standardisation. Furthermore, the conclusion is the part that an institution will consume in order to represent something as nettable or non-nettable. Ideally, an institution wants this part to be as clearly stated as possible: it would function on the basis that if certain conditions are met (such as the assumptions or qualifications) then an institution can determine the outcome based on the input of relevant factors (for example, counterparty type and product).

Encouraging lawyers to use a clear and structured format rather than nuanced text may be problematic, not least because of the potentially conflicting drivers for the law firm and the trading institution. The law firm may wish to reduce its liability by watering down the certainty of the conclusions, whereas the trading institution is likely to prefer a clearly stated opinion on which it can rely. This is not only problematic due to the use of phrases such as "the courts are likely to", "the courts should" and "the courts would" but also because different law firms may have different approaches to managing risk and so the use of such phrases across different law firms may not be consistent. Ideally, the legal opinion in structured form would support levels of certainty and nuance but do so in an objective and standardised manner that is consistent across legal opinions sourced from different law firms. This might for example be achieved via selection from pre-defined phrases, or by use of a numerical scale of assurance, and it will be necessary to distinguish between positive and negative assurances (e.g. "It is likely that an English court would …" and "It is likely that an English court would not …"). Lawyers may feel a numerical scale is inappropriately exact (i.e. suggesting a degree of accuracy that is neither intended not possible), but may be more amenable to selection from a small number of choices of increasing probability (e.g. "unknown", "definitely not", "possible", "more likely than not", and "definitely"). Effectively, what is proposed is the use of a controlled natural language.[34]

---

[33] McCann Fitzgerald (2020), 'Enforceability of the Close-out Netting Provisions of the 1992 ISDA Master Agreements and the 2002 ISDA Master Agreement', Appendix 10.

A possible approach is to construct sentences by selecting likelihood, object, verb and predicate from dropdown lists. For example, likelihood could be selected from {unknown whether, definitely not the case that, possible that, more likely than not that, definitely the case that}, object might be selected from {transactions, collateral, enforcement of close-out netting}, verb might be chosen from {is, is not, will be, will not be, can be, cannot be} and finally the predicate could be chosen from {cherry-picked, enforceable, stayed}. In this example, if each sentence starts with "It is", the following sentences can be generated:

- It is possible that transactions will be cherry-picked
- It is more likely than not that transactions will not be cherry-picked
- It is more likely than not that collateral will not be enforceable
- It is unknown whether enforcement of close-out netting can be stayed

The first two of the above example sentences are not alternatives and they may both appear in the conclusions. Law firms should be encouraged to be as specific as possible, for example by using both the "will" and "will not" forms of a sentence. We note also that the phrase "more likely than not" is preferable to the alternative phrase "probable" since the former can be more precisely mapped to numerical probabilities.

### 4.2. Netting determination

Whilst the controlled natural language for legal opinions will provide greater consistency and clarity of expression, it is the trading institution that must apply its own interpretation of the conclusions (according to its own overall appetite for risk) and apply different weightings to the different aspects of the legal opinion, in relation to the facts of each master agreement, and derive an overall positive or negative outcome relating to netting.

With smart close-out netting, the netting determination is achieved within a largely data-driven framework as follows:

In the context of each master agreement, the institution studies the discussion part of the legal opinion, together with the assumptions, qualifications and conclusions, and derives (and stores) its assessment of the legal reasoning as it relates to the master agreement. This part requires human assessment and is not driven by an algorithm.

The conclusions are assumed to be expressed in a structured form and are converted into data in an entirely mechanistic manner:

- For each relevant risk factor, the institution will select from the legal opinion all structured sentences that relate to that risk factor.

---

[34] *Kuhn surveys existing English-based controlled natural languages (CNLs), including 100 CNLs from 1930 to 2014: Kuhn, T. (2014), 'A Survey and Classification of Controlled Natural Languages', Association for Computational Linguistics.*

- The institution will then apply a custom mapping to convert "likelihood" into numerical data – perhaps to generate a single probability, or a range of probabilities. This mapping will be defined by the institution's own appetite for risk.
- The institution applies a pre-assigned weighting to each relevant risk factor (according to the institution's risk tolerance levels) and combines them all to derive an overall expression of risk.

Exactly how the overall expression of risk is achieved will be different for each trading institution – it could for example use a simple linear model that derives the sum of each risk factor (such as cherry-picking by a liquidator, or collateral enforceability) multiplied by a weight. However, a more nuanced polynomial approach may be required if the risk factors are not deemed to be entirely independent.

The overall netting determination is derived from the results of 1 and 2 above.

As an example of the approach that might be taken during netting determination, the risk that transactions will be cherry-picked by an insolvency practitioner can be assessed by selecting from the conclusion all sentences whose object is "transactions" and whose predicate is "cherry-picked". Using the examples given above, this would yield two sentences. An institution might map the likelihood to a range of percentage probabilities as follows (note that different institutions might apply different mappings according to their appetite for risk):

- Unknown whether = 0% - 100%
- Definitely not the case that = 0% - 0%
- Possible that = 1% - 64%
- More likely than not that = 51% - 100%
- Definitely the case that = 100% - 100%

The first sentence ("It is possible that transactions will be cherry-picked") therefore gives a 1% - 64% probability of cherry-picking and the second sentence ("It is more likely than not that transactions will not be cherry picked") gives a 51%-100% probability of not cherry-picking (which equates to a 0%-49% probability of cherry-picking). The intersection of the two ranges gives a probability range of 1% - 49% of cherry-picking.

This would be repeated for each risk factor (e.g. "collateral will not be enforceable", "enforcement of close-out netting can be stayed", …) to produce a probability range for each risk factor. Each risk factor would be multiplied by a weighting specified by the institution and all of them combined (perhaps as a linear sum, or perhaps with a more complex expression) to create a single overall expression of risk.

By utilising numerical risk assessments and institutional-specific weightings in the framework, this provides a representation of subjective elements (the legal opinion) to be utilised in an objective framework, whilst also allowing individual institutions to provide their view on the risk based on their particular factual circumstances and risk tolerances. These individual circumstances and tolerances may include elements such as political risk for certain jurisdictions and operational risk. [35] [36] This too can be represented in an objective framework. Two Saudi Arabian banks, for example, will likely not need to include a weighting for political risk when dealing with each other. Conversely, English institutions may include a weighting for political risk when dealing with a Saudi Arabian bank and each institution's weighting will differ.

---

[35] Political risk can be defined as the risk created by a particular jurisdiction's political framework and approach to financial trading.

[36] Operational risk is the risk of direct or indirect loss resulting from inadequate or failed internal processes, people and systems or from external events (Basel Committee on Banking Supervision (2001), *Consultative Document: Operational Risk* - https://www.bis.org/publ/bcbsca07.pdf).

## 5. Interaction with other initiatives for standardisation

### 5.1. ISDA Clause Library

We have previously explained how the ISDA Master Agreement assists in reducing the variation of possibilities for an OTC derivatives contract and establishes a broad close-out netting framework agreement for trades. This in turn reduces the number of bespoke opinions a firm may require. Nonetheless, the ability for parties to make elections (such as governing law or submission to jurisdiction) and add bespoke language in the ISDA Schedule can result in the following two scenarios:

1. similar elections expressed differently requiring consideration in a legal opinion. This additional analysis is vulnerable to misunderstanding of the intended semantics, and may have unintended consequences for close-out netting; and / or
2. an entirely new opinion being required due to significant changes not contemplated by the original opinion or changes which oppose an assumption. This additional analysis will be presented in a bespoke ancillary opinion usually, necessitating analysis across multiple opinions. Clearly this adds to the complexity of structuring the opinion(s).

This variation was one of the core reasons for ISDA to launch the ISDA Clause Taxonomy and Library Project in 2018.[37] One major objective of this work is to identify provisions within the Schedule to the ISDA Master Agreement that may benefit from further standardisation. The project has identified opportunities for creating standard form drafting options in respect of these provisions. In doing so, the project indirectly assists with the standardisation of legal opinions and consequently the derivative lifecycle proposed by the ISDA Common Domain Model (ISDA CDM).[38] For our two scenarios the impact is:

1. The legal opinion now contains links to specifically noted clauses in the Clause Library. The opinion will illustrate the legal outcomes on the variation of the business outcomes relevant to each clause contained in the Clause library. Institutions can then use the Clause Library to select their preference whilst understanding the legal risk. As a result, the difficulty in producing the conclusion part of the legal opinion for different scenarios is reduced, there is less risk of errors and less analysis is required to be checked by an institution.
2. While significant changes can still occur, by standardizing how these changes are made, the link between differing opinions can be more uniform and more easily expressed. This furthers the inter-operability of the logic proposed for the conclusion.

As the Clause Library project becomes more developed it is foreseeable that it will enable institutions and law firms to separate out the key business outcomes defined within the Master Agreement, which would be unlikely to change, and those which are less standard but of equal importance, usually elections found in the Schedule. The Clause Library could, for example, help parties assess whether a key business outcome, such as the recognition of the single agreement theory embedded in Section 1(c) was applicable and could be relied upon. This would be key to ensuring netting. An alternative business outcome would be the election of automatic early termination under Section 6(a); this could be a requirement of netting or something an institution views as useful but not essential.

---

[37] ISDA (2018), 'Launch of the ISDA Legal Technology Working Group and the ISDA Clause Library Project' (https://www.isda.org/a/DZdEE/ISDA-Clause-Library-Project-Memo.pdf).

[38] ISDA (2018), 'What is the ISDA CDM', https://www.isda.org/a/z8AEE/ISDA-CDM-Factsheet.pdf.

In time, this could assist in the production of legal opinions. Currently, the majority of legal opinions will have a standard assumption that there 'are no other material changes' without clearly defining a material change. With the Clause Library clearly defining between these key outcomes and elective outcomes, law firms can more easily define what they mean by 'no other material changes' by reference to the Clause Library, since the Clause Library represents a common vernacular.

In summary, the aim of the Clause Library Project is to reduce the variation in language within the agreement; this contributes to the ability to further standardize and automate the framework of opinion production by reducing the unseen variables. Further, when the Clause Library Project is seen in tandem with ISDA CDM and its data driven methodology the potential to assist in the derivatives lifecycle is clear.

### 5.2. ISDA CDM

Common "trigger" events in the legal opinion lifecycle, such as a new trade type, a change in the positive and negative exposures of the transactions related to the master agreement which leads to close out netting being more relevant, a change in the trades governed by the master agreement or a new legal opinion, are currently represented differently by different institutions. Indeed, it is possible for institutions to have representations for events that differ between internal systems. The ISDA CDM provides a common representation of data structures and processes across a wide range of financial products.[39] When fully implemented, the ISDA CDM will be able to represent trade events using a uniform data definition and will support significant interoperability between institutions and platforms. Institutions will therefore be able to improve close-out netting operations by integrating their legal opinion and legal agreement management system with the ISDA CDM. The CDM data is machine-readable and the potential ability to recognise automatically a change in master agreement terms or a change in insolvency law could drastically reduce the cost and effort of processing data across institutions and within them when it comes to managing a derivatives trade lifecycle.

The ISDA CDM offers an ability to provide systematic mapping of data for a derivatives trade. Key elements of data lineage can be built in at various levels of an institutions systems, data about the type and function of the event, alongside time stamping, produce clear data evidence of how and when the event took place. Vitally, this information is standard and uniform amongst institutions meaning it can easily be shared and understood when needed.

What may be a welcome side effect with ISDA CDM is that as firms begin to align their data more coherently, the data required for producing a legal opinion will become more aligned. In turn, different law firms, using the structure suggested above, will be motivated to align their logic to each other more closely as lifecycle events and relevant data will be presented more uniformly.

---

[39] Clack, C.D. (2018), 'Design discussion on the ISDA Common Domain Model', Journal of Digital Banking 3 (2), 165-187.

## 6. Summary and conclusion

Smart Close-out Netting aims to standardise and automate specific operational aspects of the legal and regulatory processes of close-out netting for prudentially regulated financial institutions. We have provided a review of close-out netting and its benefits for prudentially regulated trading institutions, analysed the process of using legal opinions for netting determination and the typical structure of those legal opinions, assessed the opportunities for standardisation and automation, and given a brief assessment of some of the sources of inefficiency in current practice. Smart Close-out Netting has been introduced in outline as a combination of two proposals: (i) the use of a standardised controlled natural language in legal opinions provided by law firms (especially in the conclusions of those opinions), and (ii) the use of a numerical framework during netting determination to convert the controlled natural language into data and to capture risk assessment into a standard algorithm that is consistently applied within a trading institution. In this way, Smart Close-out Netting aims to reduce complexity and inefficiency via standardisation and automation. We expect to report publicly on the details of the controlled natural language and framework implementation in a subsequent article. Finally, we have discussed how Smart Close-out Netting interacts with two other current industry initiatives for standardisation.

Further research is required on the feasibility of providing more structure to the discussion part of a legal opinion. At first sight, it appears as though this may not be cost effective, but this has not yet been securely established. It is possible that with further work substantial additional benefit may accrue from the use of a more formal logic, standardised across law firms. We hope the topics raised in this paper will stimulate discussion and look forward to receiving feedback.